\documentclass[twocolumn,showpacs,pra]{revtex4}
\pdfoutput=1
\usepackage{amssymb}
\usepackage{subfigure}
\usepackage{graphicx}
\usepackage{amsmath}
\begin{document}
\newcommand{\bra}[1]{\langle #1|}
\newcommand{\ket}[1]{|#1\rangle}
\newcommand{\braket}[2]{\langle #1|#2\rangle}
\newcommand{\sgn}{\mathrm{sgn}}

\title{Vortex Nucleation in Bose-Einstein Condensates Subject to Light Induced Effective Magnetic Fields}
\author{D. R. Murray}
\affiliation{Dept. of Physics, SUPA, University of Strathclyde, Glasgow G4 0NG, UK}
\author{P. \"Ohberg}
\affiliation{Dept. of Physics, SUPA, Heriot-Watt University, Edinburgh EH14 4AS, UK}
\author{Dami\'{a} Gomila}
\affiliation{Instituto de F\'{i}sica Interdisciplinar y Sistemas Complejos (IFISC,CSIC-UIB), Campus Universitat Illes Balears, E-07122 Palma de Mallorca, Spain}
\author{Stephen M. Barnett}
\affiliation{Dept. of Physics, SUPA, University of Strathclyde, Glasgow G4 0NG, UK}
\date{\today}

\begin{abstract}
We numerically simulate vortex nucleation in a BEC subject to an effective magnetic field.  The effective magnetic field is generated from the interplay between light with a non-trivial phase structure and the BEC, and can be shaped and controlled by appropriate modifications to the phase and intensity of the light.  We demonstrate that the nucleation of vortices is seeded by instabilities in surface excitations which are coupled to by an asymmetric trapping potential (similar to the case of condensates subject to mechanical rotation) and show that this picture also holds when the applied effective magnetic field is not homogeneous.  The eventual configuration of vortices in the cloud depends on the geometry of the applied field.
\end{abstract}
\maketitle

\section{Introduction}
A spectacular property of superfluid systems is their ability to support quantized vortices.  These can appear as flux lines in superconductors to which a sufficiently strong magnetic field has been applied.  Alternatively, in the case of neutral superfluids subject to sufficiently fast external rotation, they exist as lines of vanishing condensate density around which the velocity field flow is quantized.  Both scenarios are closely interlinked, because the equations describing a rotating superfluid, when studied in the rotating frame, mimic those of a charged superfluid (a superconductor) in a magnetic field, with the Coriolis force playing the role of the Lorentz force.

The dilute gas BEC is an extremely useful tool for probing the underlying physics of superfluid phenomena because in experiment its parameters are typically much easier to manipulate than for other condensed matter systems.  Moreover, because of its diluteness, it is considerably more amenable to theoretical treatment which more accurately reflects the experimental reality.  A good example is in the studies of condensates rotated by an anisotropic potential, where the critical velocity for vortex nucleation has been found to coincide with dynamical instabilities in the surface mode excitations \cite{SinhaCastin,vortdalibard}.  Yet, the precise mechanism for the nucleation of vortices is still debated, as are questions about the importance of thermal effects and how, if at all, the mechanism for vortex nucleation due to a rotating trap is related to generation of vortices by localized stirring \cite{PhysRevLett.87.210402}.  At the same time, the prospect of simulating effects related to charged particles in magnetic fields, means that the rotating BEC continues to receive a lot of attention \cite{PhysRevLett.87.060403,PhysRevLett.87.120405,PhysRevLett.90.140402}.

However, rotating a condensate only provides access to a limited class of problems, for which the effective magnetic field is spatially homogeneous in the plane perpendicular to the rotation axis.  In addition, controlled stirring of a BEC can be a rather demanding task.  Recent proposals to create effective magnetic fields in a more direct way open the door for more wide-ranging studies into the interaction of degenerate quantum gases with effective magnetic fields.  One method involves using lasers to alter the state-dependent tunneling amplitudes of atoms in an optical lattice to simulate an effective magnetic flux \cite{Ruostekoski02,zollerBeff,mueller}.  Another, considered here, exploits the interaction of $\Lambda$-type three-level atoms with two laser beams possessing relative orbital angular momentum in an electronically induced transparency (EIT) configuration, such that the vector potential shows up in the effective equation of motion \cite{ferfield,magfield05} for the atoms, which sit in a non-degenerate eigenstate of the laser-atom interaction.  An advantage of this method is that the vector potential, and consequently the effective magnetic field, can be shaped and controlled by appropriate modifications of the phase and intensity of the incident light.

In this paper, we study by numerical simulation the influence of both homogeneous and inhomogeneous effective magnetic fields on the dynamics of a harmonically trapped  Bose-Einstein Condensate, and observe vortex nucleation for critical parameter values.  The exact dynamics are specific to the geometry of the trapping potential and effective magnetic field, but the the existence of unstable modes in the spectrum of elementary excitations as a precursor to vortex nucleation is a universal feature for all cases considered.  Recent advances in light beam shaping technology, using for instance spatial light modulators, mean that all the potentials we consider can realistically be created in the laboratory \cite{JohannesNature}.

The paper is organized as follows: in section \ref{The Model} we outline the method for creating the effective magnetic field.  In section \ref{Sec Elliptic} we study the dynamics when a small ellipticity is introduced into the trapping potential, drawing attention to key differences between the homogeneous and inhomogeneous effective magnetic fields.  In section \ref{Sec Higher} we observe vortex nucleation traps of higher than two-fold symmetry, where the trap symmetry couples to higher order surface modes.  Finally, in section \ref{Conclusions} we summarize the main results.

\section{The Model}
\label{The Model}

\begin{figure}[tbp]
\center{
\includegraphics[width=7cm]{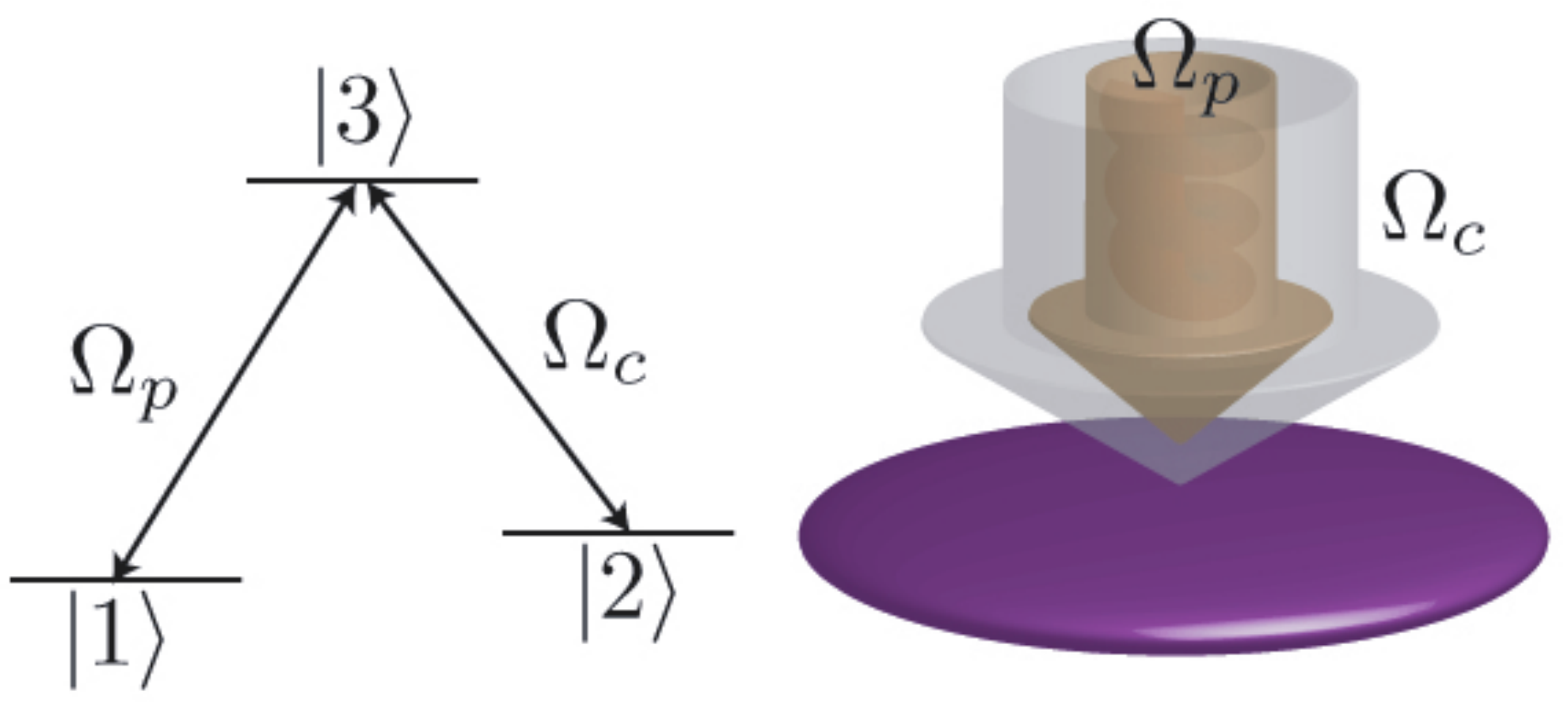}
\caption {Method for creating effective magnetic fields in degenerate atomic gases.  On the left is the level scheme for the $\Lambda$-type atoms interacting with the resonant probe beam $\Omega_{p}$ and control beam $\Omega_{c}$.  On the right is a schematic representation of the experimental setup with the two light beams incident on the cloud of atoms.  The effective magnetic field is generated if there is relative angular momentum between the beams.  This will occur, for example, if the probe field is of the form $\Omega_{p} \sim \rm{e}^{i\ell\phi}$, where each probe photon carries an orbital angular momentum $\hbar\ell$ along the propagation axis $z$, and $\Omega_{c}$ is independent of the azimuthal angle. }
\label{figlevels}}
\end{figure}
Let us briefly describe the key ingredients for creating an effective magnetic field using light with orbital angular momentum (a  full account can be found in \cite{magfield05}).  We consider a typical Electromagnetically Induced Transparency (EIT) setup with $\Lambda$-type three level atoms, characterized by two hyperfine ground levels $\ket{1}$ and $\ket{2}$ and an electronically excited level $\ket{3}$, interacting with resonant laser beams which are both propagating in the $z$-direction (see figure 1).  The atoms are assumed to be trapped in the $x-y$ plane with motion in the z-direction effectively frozen out ($\omega_{z}\gg \omega_{\perp}$) \footnote{This ensures that the effective magnetic field depends only on being able to shape the transverse laser fields}.  The probe beam, with coupling strength $\Omega_{p}$, and control beam, with coupling strength $\Omega_{c}$ drive the transitions $\ket{1} \rightarrow \ket{2}$ and $\ket{1} \rightarrow \ket{3}$ respectively.  These absorption paths interfere destructively to suppress transitions to level $\ket{3}$ and drive the atoms to the dark state $\ket{D}=\frac{\ket{1}-\zeta \ket{2}}{\sqrt{1+|\zeta|^2}}$, where $\zeta=\frac{\Omega_{p}}{\Omega_{c}}=\left |\frac{\Omega_{p}}{\Omega_{c}}\right | \mathrm{e}^{iS}$ and $S$ is the relative phase between the probe and control beam.  Provided that gradient of the phase $S$ is non zero, then, because of the spatial dependence of the dark state, two additional geometric potentials appear in the effective equation for the wavefunction of the dark state atoms.  One of these acts as an effective vector potential $\mathbf{A}$ (or Berry connection \cite{Berry84}) and the other is a scalar potential.

One way to achieve a non-zero phase gradient in the ratio of the control and probe beam coupling strengths is to use light with orbital angular momentum.  A Laguerre-Gaussian beam, for example, has transverse field  $u_{p \ell}(r,\phi) \propto r^{|\ell|}L^{|\ell|}_{p}\left(r^2\right)\exp(-r^2)\exp(i \ell \phi)$, where $L^{\ell}_{p}$ is the associated Laguerre polynomial, such that each photon carries angular momentum $\ell \hbar$ around the z-axis \cite{Allen92}.  If one or both of the control and probe beam are of this type we may consider situations where the dimensionless parameter $\zeta$ is of the form
\begin{equation}
\label{zeta}
\zeta=\sqrt{\alpha_0}\left(\frac{r}{R}\right)^{\frac{\nu+1}{2}}\exp(i \ell \phi),
\end{equation}
where the parameter $\alpha_{0}$ is the intensity ratio at $r=R$.  The mean field equation for the wavefunction of the dark state atoms is then
\begin{equation}
i\hbar\frac{\partial\Psi}{\partial{t}}=\frac{1}{2M}(i\hbar\nabla+\mathbf{A})^{2}\Psi+(U+\phi)\Psi+g \left| \Psi\right| ^{2}\Psi,
\label{TDGP1}
\end{equation}
where, if $|\zeta|^2 << 1$, the induced vector potential is well approximated by \cite{excite07}
\begin{equation}
\mathbf{A}=-\frac{\hbar \ell}{R}\alpha_{0}\left(\frac{r}{R}\right)^{\nu} \mathrm{e}_{\phi},
\label{approxA1}
\end{equation}
which corresponds to an effective magnetic field
\begin{equation}
\mathbf{B}=-\frac{\hbar\alpha_{0}\ell(\nu+1)}{R^2}\left(\frac{r}{R}\right)^{\nu-1}\mathrm{e}_{z}.
\label{approxB}
\end{equation}

In the $|\zeta|^2 << 1$ limit, the scalar potential is
\begin{equation}
\phi=\frac{\hbar^2}{2MR^2}{\left[\ell^2+\left(\frac{\nu+1}{2}\right)^2\right]}\left(\left(\frac{r}{R}\right)^{\nu-1}-2\alpha_{0}\left(\frac{r}{R}\right)^{2\nu}\right),
\label{exactV}
\end{equation}
with the external trapping potential $U$ for the dark state atoms given by
\begin{equation}
U=V_{1}+\alpha_{0}\left(\frac{r}{R}\right)^{\nu+1}V_{2},
\label{Vdark}
\end{equation}
where $V_{j}$ is the trapping potential for the atoms in hyperfine state $j$ ($j=1,2$).  We assume that the (two-dimensional) coupling constant $g$ characterizing the strength of atom-atom interactions within and between levels $\ket{1}$ and $\ket{2}$ is constant throughout the cloud which is valid if the inter- and intra-species s-wave scattering lengths are approximately equal or if $|\zeta|<<1$ \cite{excite07}.

A consequence of choosing light beams of this form is that the effective vector potential points in the $\rm{e}_{\phi}$ direction, inducing a rotational motion.  To elucidate the connection with rotating condensates, we use the fact that $\nabla \cdot \mathbf{A} =0$ to re-write Eq. (\ref{TDGP1}) as
\begin{equation}
i\hbar\frac{\partial\Psi}{\partial{t}}=\left(-\frac{\hbar^2}{2M}\nabla^2+\tilde{V}+g \left| \Psi\right| ^{2} +\frac{i\hbar}{M}\mathbf{A}\cdot\nabla\right)\Psi,
\label{GProt}
\end{equation}
where $\tilde{V}(r)=V+\frac{|A|^2}{2M}$.  An equation of identical form to Eq. (\ref{GProt}) can be obtained by considering a BEC subject to a time-dependent trapping potential which is rotating at angular velocity $\Omega$, and studying the resultant Gross-Pitaevskii equation in the rotating frame in which the trapping potential is independent of time.  The GP equation in this frame is as Eq. (\ref{GProt}) where the effective vector potential is given by $\mathbf{A}=M\mathbf{\Omega}\times\mathbf{r}$, which would correspond to an effective magnetic field which is homogeneous in the z-direction $\mathbf{B}=2M\Omega \rm{e}_{z}$.

 The ability to shape the phase and intensity of the incident light, however, means that we are no longer restricted to considering only homogeneous effective magnetic fields, unlike the case of the condensate subject to external rotation.  Actually, we can consider Eqs. (\ref{GProt}) and (\ref{approxA1}) as analogous to a rotation of $\frac{A}{r}$.  A shearing, or \textit{differential} rotation (one at which different parts of the cloud are rotating at different angular velocities) is obtained if the vector potential $\mathbf{A}$ corresponds to an inhomogenous magnetic field.

 The possibility of studying inhomogenous effective magnetic fields raises many interesting questions, including: do we still observe vortex nucleation above a critical field strength?  Does the same mechanism for vortex nucleation - attributed to instabilities of the surface mode excitations - still hold for effective magnetic fields which are inhomogeneous?  If so what does the lattice configuration look like?

We shall address these questions in turn, using results of numerical simulations of the real-time dynamics of the condensate as determined by Eq. (\ref{GProt}), where the vector potential is of the form given by Eq. (\ref{approxA1}).  We shall consider the trapping potential as it appears in Eq. (\ref{GProt}) to be of the form:
\begin{equation}
\tilde{V}=\frac{1}{2}M\omega_{\perp}^2 r^2+ \Delta{V}_{asym.},
\label{tildeV}
\end{equation}
that is, harmonic plus a small perturbation which is not radially symmetric.  A harmonic potential can be achieved by a judicious choice of the external trapping potential (Eq. \ref{Vdark}) to counteract the additional scalar potentials due to the light-atom interaction.  Our motivation for considering potentials of this (predominately harmonic) form is two-fold.  First, this is of the same form as the rotating frame potential of atoms in a rotating elliptical harmonic trap, allowing for a clear comparison.  Also, because we are primarily interested in vortex nucleation, it is instructive to eliminate the scalar potential which can preclude vortex nucleation in inhomogeneous magnetic fields by preventing the necessary surface mode instabilities from occurring \cite{excite07}.

To simulate the evolution of the condensate according to Eq. (\ref{GProt}), we first  re-scale length and time variables as

\begin{eqnarray}
  \nonumber \{x,y,R\}&\rightarrow& \left(\frac{2\mu}{M\omega_{\perp}^2}\right)^{\frac{1}{2}}\{x,y,R\} \\
  t &\rightarrow& \omega_{\perp}t,
  \label{dimvariables}
\end{eqnarray}
respectively, where $\mu$ is the chemical potential.  This allows us to write Eq. (\ref{GProt}) in the dimensionless form
\begin{eqnarray}
i\gamma \partial_{t}\psi=\bigg[-\gamma^2\nabla^2+x^2+y^2+\Delta{\tilde{V}}_{asym.} +\tilde{g}|\psi|^2\nonumber\\+
2i\gamma^2\frac{\alpha_{0}\ell}{R^2}\left(\frac{r}{R}\right)^{q-1}(-y\partial_{x}+x\partial_{y})\bigg]\psi,
\label{GProtdim}
\end{eqnarray}
where $\tilde{g}=\frac{g}{\mu}$ and the parameter $\gamma=\frac{\hbar\omega_{\perp}}{2\mu}\ll1$ in the Thomas-Fermi regime.  We obtain the initial state by relaxing Eq. (\ref{GProtdim}) in imaginary time to find the ground state where we constrain $\psi$ to be real, and then propagate this solution according to Eq. (\ref{GProtdim}) by an easily implemented leap-frog method \cite{numericalrecipes}, checking that the energy and the norm are preserved throughout.  This situation corresponds to a sudden switch on of the effective magnetic field at time $t=0$.

To simulate experimental imperfections and to break the unrealistic levels of symmetry \cite{parkerTURBULENCE}, we add small fluctuations ($\backsim 0.01 \%\langle\tilde{V}\rangle  \ll \mathrm{grid\;spacing}$) to the trap coordinates periodically (not more often than every $10^5$ time steps).  This does not change the dynamics qualitatively but speeds up the symmetry breaking, which would otherwise occur due to growth of numerical noise \cite{parkerTURBULENCE}.

\section{Elliptic Perturbation to the trap}
\label{Sec Elliptic}
\subsection{Homogenous Magnetic Field ($\nu=1$)}
An already well-studied problem in respect of its equivalence to rotating condensate experiments is the homogeneous effective magnetic field ($\nu=1$) applied to a condensate in a harmonic trap with a small elliptic perturbation \cite{vortdalibard,SinhaCastin,Recati2001,vortHodby}
\begin{equation}\label{eq-ellpert}
\Delta{{V}}_{asym.}=\epsilon\left(x^2-y^2\right),
\end{equation}
where $\Delta{{V}}_{asym.}$ is written in the dimensionless units (Eq. \ref{dimvariables}).  The addition of the asymmetry in the potential distorts the flow $\mathbf{J}=\frac{\hbar}{2i}\left(\psi^{*}\nabla \psi-\psi\nabla\psi^{*}\right)-\mathbf{A}|\psi|^2$, leading to shape oscillations.  Because the trap has a two-fold symmetry, it naturally couples to the quadrupole oscillations which carry $m=2$ units of angular momentum.   At a critical value of $\alpha_{0}\ell$, the frequency of the quadrupole mode tends to zero, and large amplitude oscillations take place.  The critical value of $\alpha_{0}\ell$ corresponds to an energetic instability of the quadrupole mode, occurring when the cyclotron frequency associated with the effective magnetic field matches the bare quadrupole frequency in absence of the field.  The large amplitude oscillations are dynamically unstable, eventually leading to nucleation of vortices in the bulk of the condensate, as shown in figure 2.

\begin{figure}
  \includegraphics[width=7cm,viewport=100 300 495 500 ]{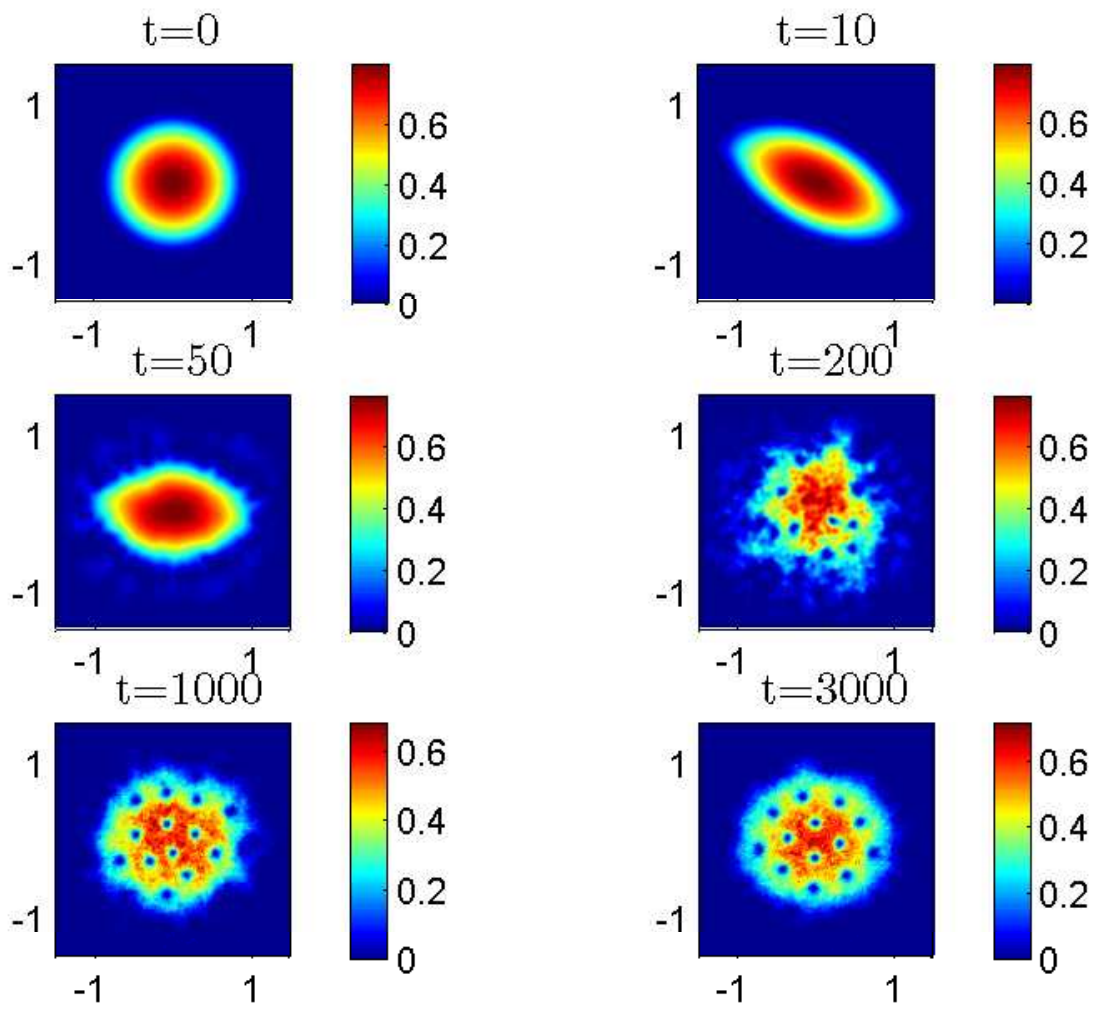}\\
  \caption{Numerical simulation of the density of a BEC in a slightly elliptic trap for $\gamma^2=0.001$, $\epsilon=0.05$ and $\alpha_{0}\ell = 23$, where the time is given in units of $1/\omega_{\perp}$.  The almost circular initial state becomes very elongated shortly after application of the effective magnetic field with cyclotron frequency resonant with the frequency of the quadrupole mode.  The dynamical instability causes fragmentation of the cloud and subsequent nucleation of vortices.  These vortices eventually settle into a stable lattice which is non-rotating in the laboratory frame.  The nucleation dynamics are in good agreement with the model presented in ref \cite{parkerTURBULENCE}, which considers a sudden switch on of rotation.} \label{fig-rpow1vort}
\end{figure}

\subsection{Inhomogeneous Magnetic Field ($\nu > 1$)}
It is natural to contemplate whether a similar situation should occur for inhomogenous effective magnetic fields, that is, do we observe vortex nucleation due to coupling of the elliptic trap to a surface mode which becomes unstable.  An obvious place to look is at the magnetic field strength $\alpha_0\ell$ for which the $m=2$ mode becomes unstable.  We have previously calculated the surface mode frequencies of a harmonically trapped condensate subject to a vector potential with $\nu \ge 2$, by linearizing around the non-vortex ground state solution of a BEC in absence of effective magnetic field \cite{excite07}.  For $q=2$ in Eqs. (\ref{approxA1})-(\ref{Vdark}) and (\ref{GProtdim}), the $m=2$ surface mode becomes unstable at $\alpha_{0}\ell\simeq 26$ for the parameter values $\gamma^2=0.001$, $R=1$ and $\tilde{g}=1$.  Simulations run with this vector potential indicate again a strong shape deformation after the effective magnetic field is switched on but rather than fragmentation followed by vortex nucleation and eventual crystallization of the vortex lattice, we observe a more dramatic shedding of density from the cloud.  The reason for these explosive dynamics can be explained by classical arguments:  the effective scalar potential `seen' by the rotating condensate is of the form
\begin{equation} V(r)=(1+\epsilon)x^2+(1-\epsilon)y^2+\Delta\tilde{V}_{asym}-(\gamma\alpha_{0}\ell)^2 r^4,
\label{unconfined}
\end{equation}
and reaches a maximum at radius $r_{max}=\frac{1+\epsilon\sin(2\theta)}{\sqrt{2}\gamma\alpha_{0}\ell}$, where $\theta$ is the polar angle.  If the chemical potential $\mu$ ($\sim 1$) exceeds the maximum potential energy of the trap $V(r_{max})$ many atoms are expelled from the cloud.  In such cases the system no longer becomes manageable for either numerical studies or actual experiments.

This phenomena is primarily a result of our choice of trapping potential (Eq. (\ref{tildeV})), chosen to allow comparisons with previous experiments.  Indeed, a similar problem exists in rotating trap experiments when the rotation frequency $\Omega$ exceeds the trapping frequency $\omega_{\perp}$ causing the cloud to explode due to centrifugal effects.  This has been overcome by adding an additional, steeper potential to keep the atoms confined.  In our case, a sufficiently steep confining potential exists due to the interaction of the atoms with the light (Eq. \ref{exactV}), but the trade-off from using this potential in its unaltered form is that the surface modes are energetically stable for all magnetic fields $\nu \ge 2$, hence no dynamic instability occurs to seed vortex nucleation.  Instead, in the next section we consider the resonant excitation of surface modes higher than $m=2$ as a route to achieving vortex nucleation in inhomogeneous effective magnetic fields.
\section{Traps of higher symmetry}
\label{Sec Higher}
Although most studies to date have considered the rotating elliptically deformed trap as a means to generate vortices due to the quadrupolar instability, it has also been demonstrated that vortex formation occurs close to predicted values when using a stirring potential which resonantly excites modes of higher multipolarity.  In the case of a BEC in a light induced effective magnetic field, a trapping potential which is of rotational symmetry j, say, should couple to surface modes of multipolarity $m=j$.  A trap with rotational symmetry higher than 2 can be achieved by applying a laser with multiple beam patterns to create the optical potential of desired symmetry \cite{ferris07}.

Let us assume the asymmetric perturbation to the radially symmetric harmonic trap (Eq. (\ref{tildeV})), written in dimensionless units, is of the form
\begin{equation}
\Delta{\tilde{V}}_{asym.}=\frac{\epsilon R^2}{\left|\cos\left(\frac{j}{4}\phi\right)\right|+\left|\sin\left(\frac{j}{4}\phi\right)\right|},
\label{eq-highsym}
\end{equation}
where $j$ gives the rotational symmetry, $\phi$ is the polar angle and $\epsilon$ is a dimensionless parameter.  We choose the parameters $\alpha_{0}\ell$ and $j$ to resonantly excite a particular mode.  In figure 3 we plot the location of the surface mode energetic instabilities for three different effective magnetic fields.  These instabilities were calculated in ref \cite{excite07}, for a BEC in a  radially symmetric harmonic trap.  The addition of the potential in Eq. (\ref{eq-highsym}) will shift the mode frequencies but not by much provided $\epsilon$ is small.

\begin{figure}
  \includegraphics[width=7cm,viewport=100 250 495 550]{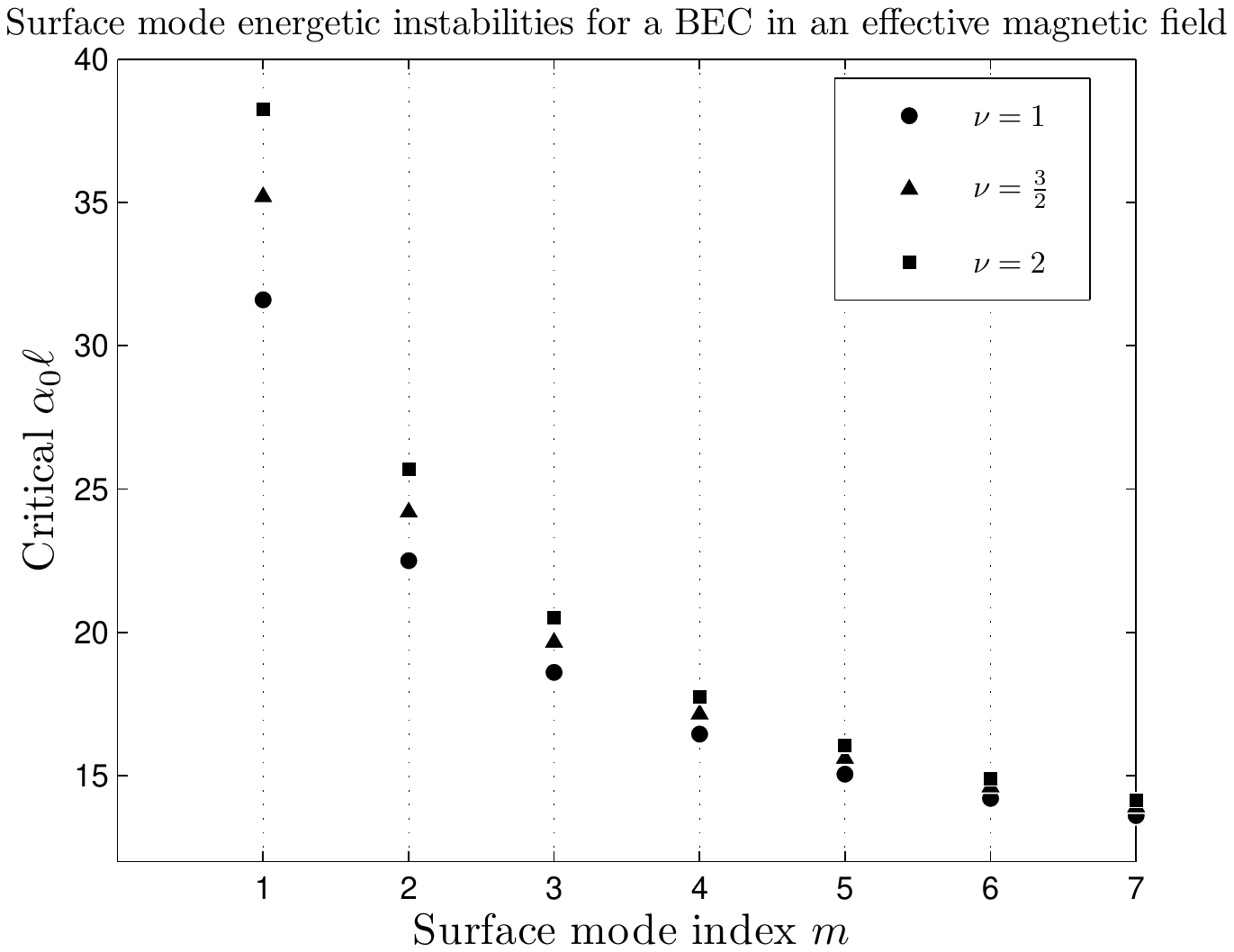}\\
  \caption{Location of the energetic instability of the surface modes $m$ for three different vector potentials of the form $\mathbf{A}\sim r^{\nu}\mathrm{e}_{\phi}$.  Above the critical field strength $\alpha_{0}\ell$, the mode excitation energy is negative, so that in the presence of dissipation the system can lower its energy by going into an 'anomolous' mode.  This instability should not be confused with the dynamical instability whose signature would be a complex excitation energy.  However, energetic instability is a prerequisite for dynamical instability.  }\label{fig-instability}
\end{figure}

In our simulations, we observe vortex nucleation for homogeneous and inhomogenous effective magnetic fields close to the instability of the mode coupled to by Eq. (\ref{eq-highsym}).  The cases we consider are when $\nu=\frac{3}{2}$ 
and $\nu=2$, which both correspond to an effective magnetic field which increases with radius $r$ (see Eq. (\ref{approxB})).  For fields of this form, it is necessary to examine vortex generation around instabilities for surface modes higher than $m=2$ to avoid expulsion of a significant proportion of atoms.  One should remember though that the linearization of the Gross-Pitaevskii equation used to calculate the excitations is less accurate for higher $m$ modes.  In figure 4, we demonstrate nucleation of vortices for the $\nu=\frac{3}{2}$ field with an asymmetric perturbation to the harmonic trap (Eq. (\ref{eq-highsym})) of rotational symmetry $j=4$, where $\alpha_{0}\ell$ is chosen to be slightly above the critical value as determined by figure 3.

The presence of energetic instability is a necessary, but insufficient criterion for vortex nucleation; it reflects only that the wave-function is no longer a local minimum of the energy.  The nucleation of vortices is due to the dynamical instability, whose signature is a complex excitation energy, allowing small amplitude perturbations to grow exponentially in time causing a significant change to the initial state.  Consistent with this picture, our studies indicate the existence of an upper as well as lower bound of $\alpha_{0}\ell$  for vortex nucleation to occur.  There is no upper bound for the energetic instability, and so the nucleation of vortices must correspond to a window of dynamical instability.  However, because energetic instability implies dynamical stability \cite{jackson:053617} for the conditions outlined in this paper, it is still a prerequisite for the vortex nucleation.  Moreover, the energetic stability drastically suppresses the energy barrier which otherwise prevents vortices entering into the cloud \cite{kramer02}.

\begin{figure}[htpb]
  \includegraphics[width=7cm,viewport=50 250 545 590]{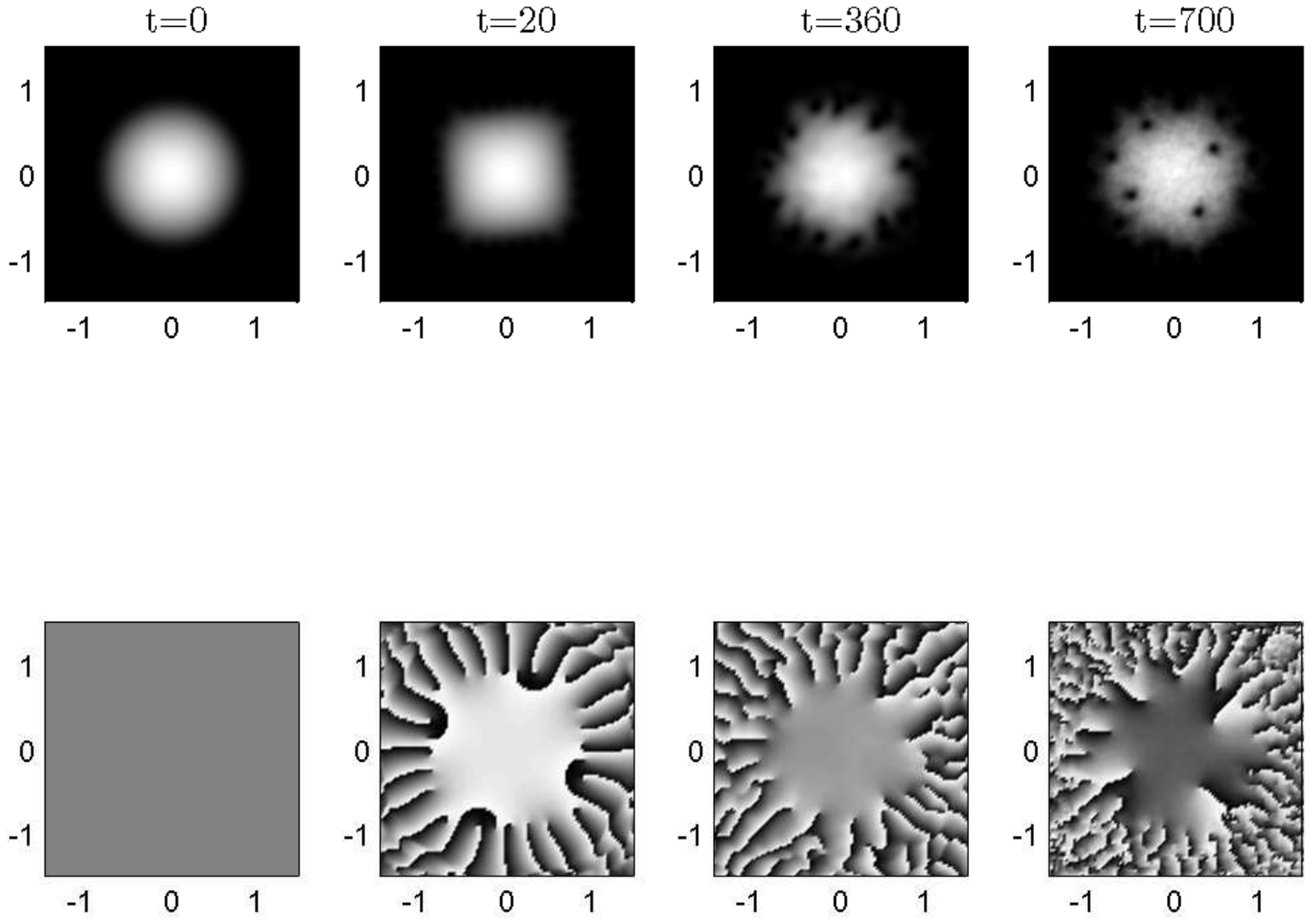}\\
  \caption{Snapshots of the Density (top) and phase (bottom) for an inhomogeneous effective magnetic field with $\nu=\frac{3}{2}$, $\alpha_{0}\ell=18.2$, $j=4$, where the time $t/\omega_{\perp}=0,20,360,700$ increasing from left to right.  The $m=4$ octopole surface mode is resonantly excited, and vortex nucleation is enabled by a dynamical instability.   }\label{fig-instability}
\end{figure}

Similar to studies of refs \cite{kramer02,vortHodby,martikainen03}, we note that the width of the unstable region increases with $\epsilon$, the anisotropy of the trap perturbation (Eq.(\ref{eq-highsym})).  For the parameters $\nu=2$ and $j=5$, we observe vortex nucleation within in the range $\alpha_{0}\ell= 15.25 - 16.25$ for $\epsilon=0.05$ and $\alpha_{0}\ell=15.0-17.0$ for $\epsilon=0.1$.

Let us also briefly mention variation in the structure of the vortex lattice as a function of the geometry of the effective magnetic field.  As is well known, the condensate in a homogenous effective magnetic field with a large number of vortices favours an Abrikosov type lattice \cite{Tilley}.  Even with a small number of vortices, we see that the lattice approaches a regular homogenous structure in figure 2 for $\nu=1$.
\begin{figure}[pb]
\centerline{
\subfigure[]{\includegraphics[width=0.9in,viewport=100 300 495 500]{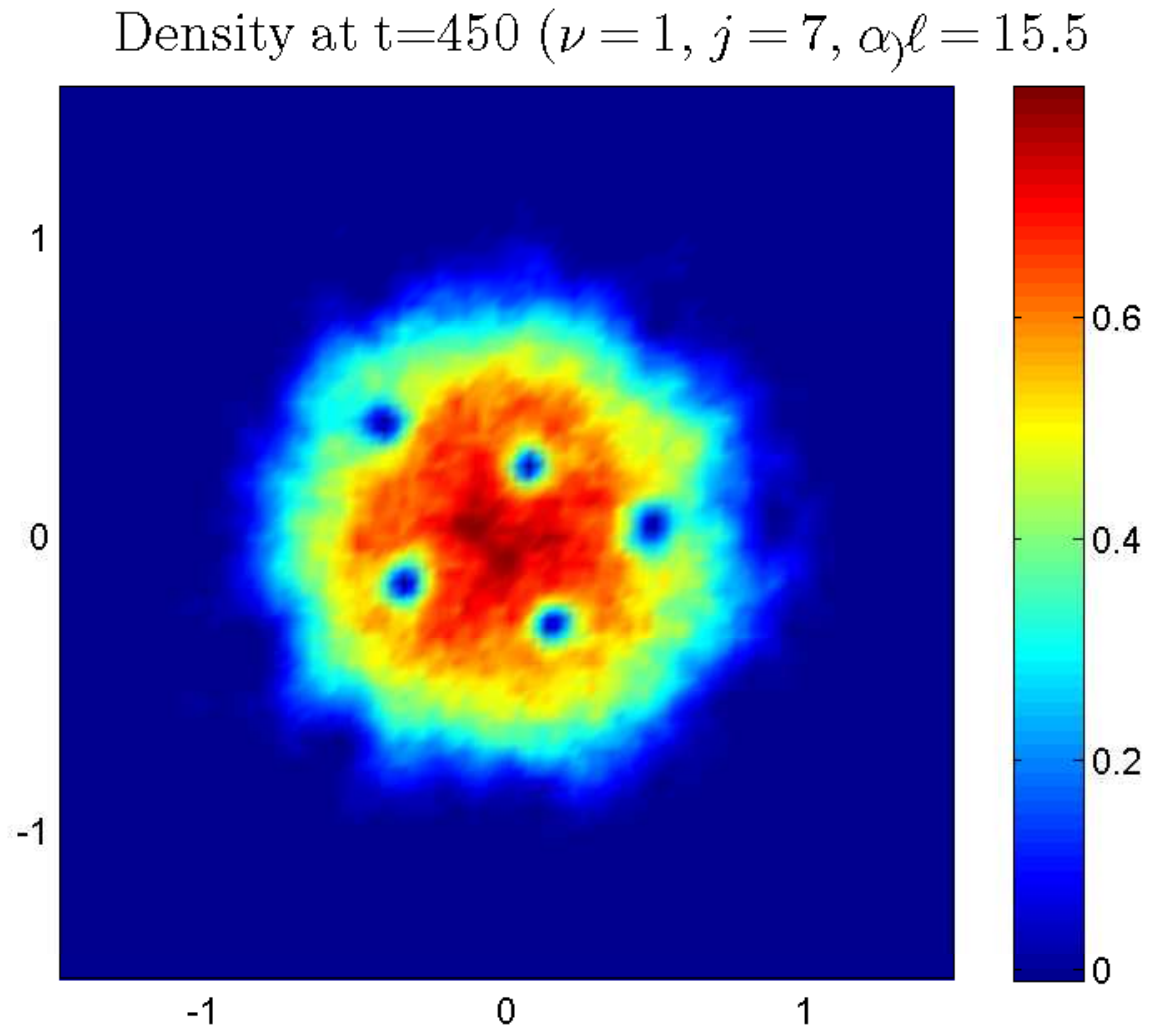}
\label{sfi:r1j7t450}}
\hfil
\subfigure[]{\includegraphics[width=0.9in,viewport=100 300 495 500]{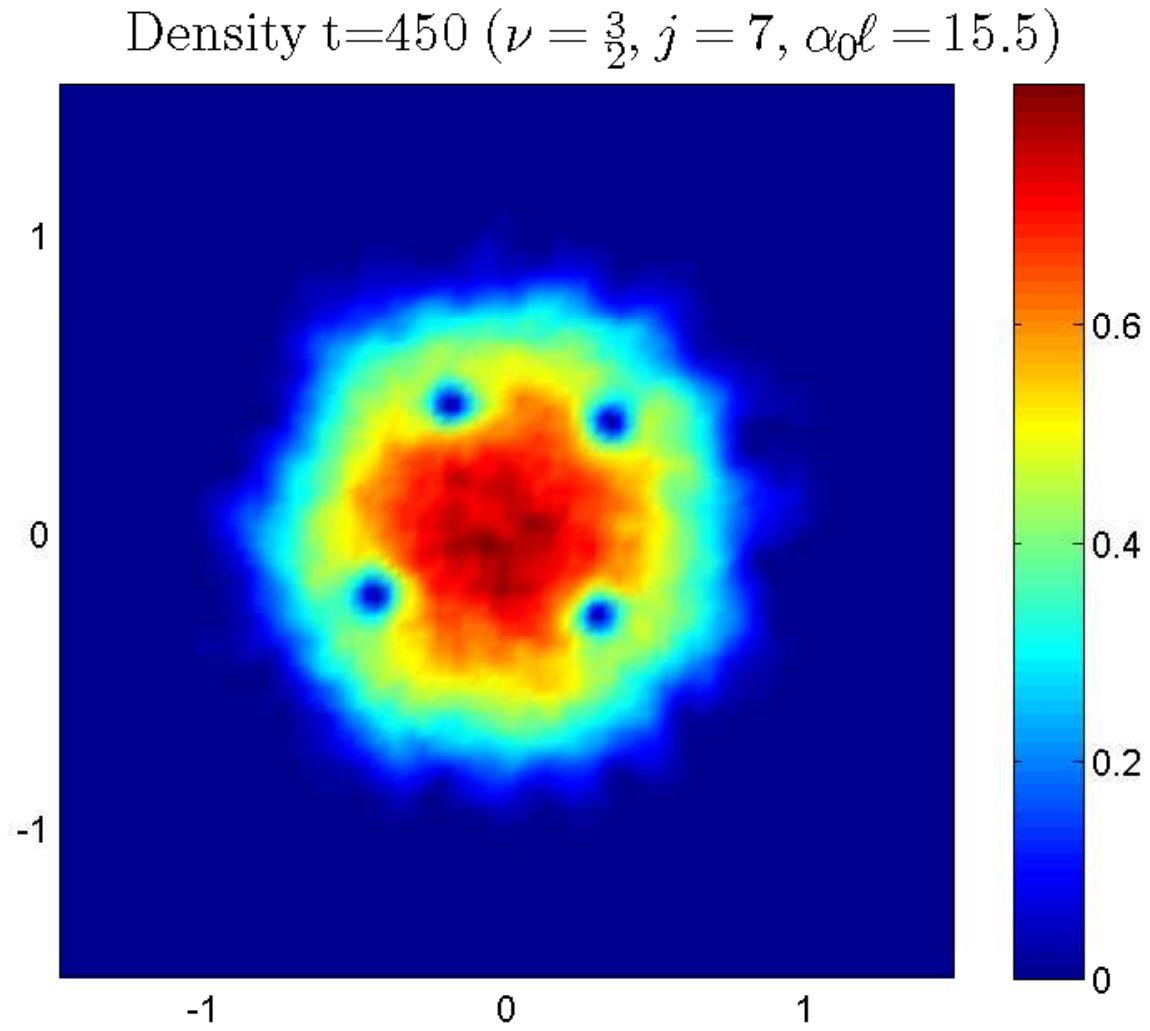}
\label{sfi:r3o2j7t450}}
\hfil
\subfigure[]{\includegraphics[width=0.9in,viewport=100 300 495 500]{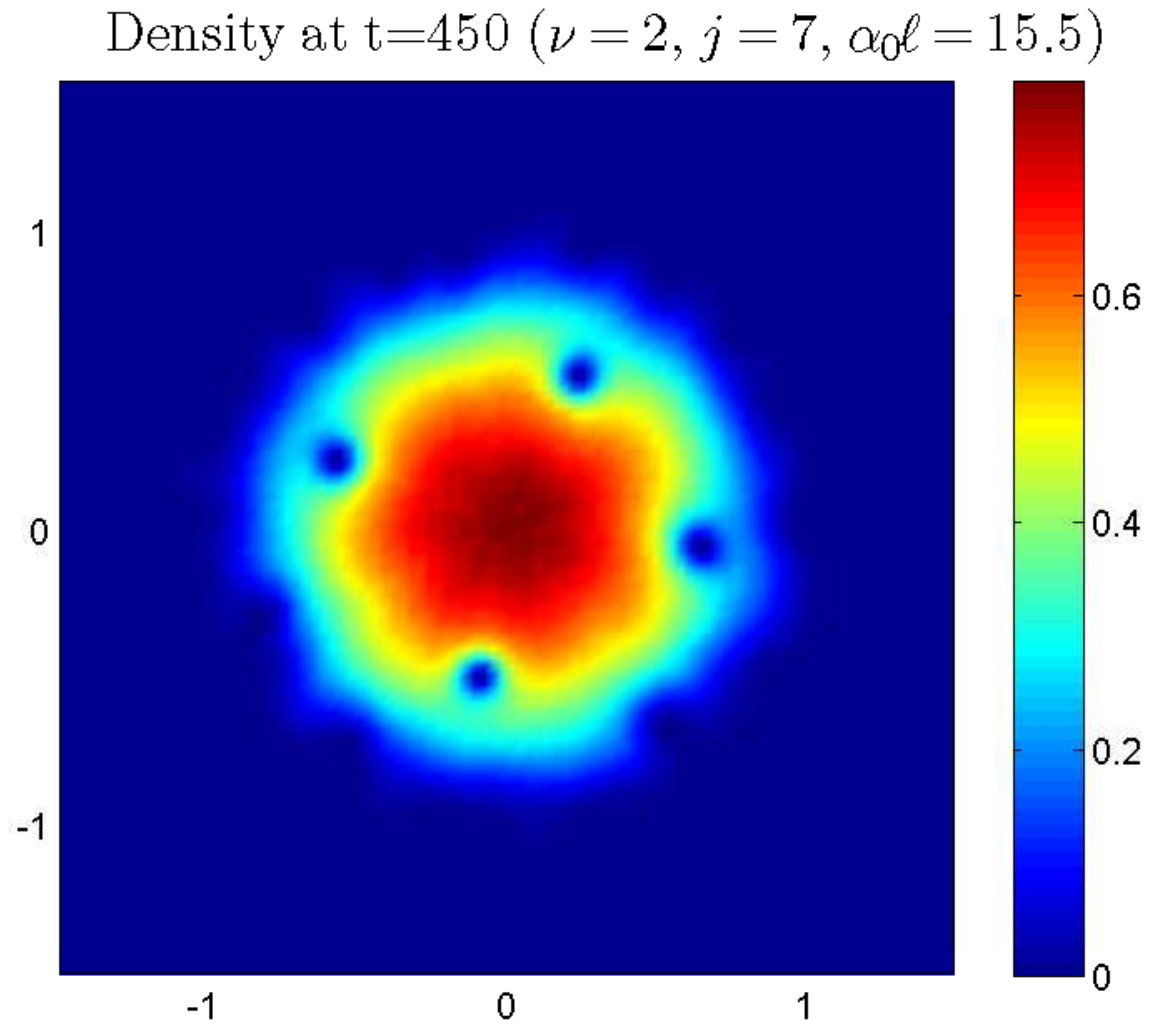}
\label{sfi:r2j7t450}}
}
\caption{Density at t=450 for BEC in an asymmetric trap subject to the effective magnetic field with (a) $\nu=1$ (homogeneous), (b) $\nu=\frac{3}{2}$ and (c) $\nu=2$.  The parameters used in (a)-(c) were $\epsilon=0.1$, $j=7$, $\alpha_{0}\ell=15.5$ and $\gamma^2=0.01$.  The nucleation of vortices is driven by resonant excitation of the $m=7$ mode, for which the critical $\alpha_0\ell$ is just below 15 for each of magnetic fields considered here (see figure 3).}
\label{fi:lattice}
\end{figure}
For inhomogeneous effective magnetic fields with $\nu > 1$, in the examples studied we note that the vortices tend to be concentrated toward the edges of the cloud where the effective magnetic field is stronger, an effect which is enhanced as $\nu$ is increased (see figure 5).  Some additional insight into this effect can be gained by using the concept of diffused vorticity \cite{feynvort}, which is applicable if the there are a large number of vortices in the cloud, and consequently a dense vortex lattice (much more so than in the examples shown here). In these conditions, the course grained average of the velocity field $v_{cg}$ associated with the lattice will be a smooth function of the radius $r$.  A dense regular array of equally spaced vortices would thus mimic solid body rotation.  Approximating the wavefunction as $\psi=\sqrt{\rho}\mathrm{e}^{iS}$, where $\rho$ and $S$ are coarse-grained averages of the density and phase, respectively, such that $v_{cg}=\frac{\hbar}{M}\nabla S$, we obtain an expression for the energy,
\begin{eqnarray}
E=\int d\mathbf{r} \bigg[\frac{\hbar^2}{2M}\nabla \rho \cdot \nabla \rho + \frac{M}{2}\left(\mathbf{v}_{cg}-\frac{\mathbf{A}}{M}\right)^2\rho \nonumber\\+V\rho+\frac{g}{2}\rho-\mu\rho\bigg].
\label{eq-EN}
\end{eqnarray}
Minimizing with respect to $\mathbf{v}_{cg}$ we find
\begin{equation}
\mathbf{v}_{cg}=\frac{\mathbf{A}}{M},
\label{Eq-vcg}
\end{equation}
independent of the local density $\rho(r)$, and the average vorticity is given by
\begin{equation}
\nabla \times \mathbf{v}_{cg} = \frac{\mathbf{B}}{M}.
\label{Eq-vorticity}
\end{equation}
Thus, it is energetically favourable for the induced vorticity to mimic the effective magnetic field, implying that for a field of the form $B\propto r^\nu$ ($\nu>1$) there should be a higher concentration of vortices at a larger radius, as we indeed see in figure 5.  Of course, this analysis takes no account of the vortex-vortex interactions (repulsive for vortices of the same `charge'), which become more important as the vortex density is increased and will tend to restore regularity to the lattice.

\section{Conclusions}
\label{Conclusions}

We have demonstrated vortex nucleation in BECs interacting with light which has a non-trivial phase structure, where the light-BEC interaction plays the role of an effective magnetic field acting on the atoms.  The field can be shaped and controlled by appropriate modifications to the light phase and intensity, allowing us to also examine the role of inhomogenous magnetic fields.

 In the examples considered, we find that the nucleation of vortices is crucially dependent on instabilities in the spectrum of elementary excitations, as is known to be the case for condensates subject to external rotation. The form of the magnetic fields considered here dictates that the dynamics are primarily determined by the surface excitations.  It is interesting to speculate whether investigation of vortex generation due to more localized effective magnetic fields could bridge a connection in the theory between experiments where the condensate is stirred by an optical `spoon' to create vortices \cite{PhysRevLett.87.210402} as opposed to rotating bucket type configurations \cite{vortdalibard}.

The use of light-induced effective magnetic fields provides a useful means of rotating a condensate without mechanical stirring.  Manipulations of the trapping potential and the incident light allow a high level of control over the vortex dynamics, including the number of vortices in the cloud and the structure of the lattice.

\begin{acknowledgements}
This work was supported by the United Kingdom EPSRC.  S.M.B thanks the Royal Society and the Wolfson Foundation. D.G. acknowledges financial support from MEC (Spain, Grant No. TEC2006-10009) and Govern Balear (Grant No. PROGECIB-5A).
\end{acknowledgements}


\bibliographystyle{apsrev}

\end{document}